\documentclass[aps,pre,twocolumn,superscriptaddress,floatfix]{revtex4}
\usepackage{graphicx}
\usepackage[dvips,unicode,colorlinks,linkcolor=blue,citecolor=blue,urlcolor=blue]{hyperref}
\begin{document}
\title{Heat conduction in diatomic chains with correlated disorder}

\author{A. V. Savin}
\affiliation{Semenov Institute of Chemical Physics, Russian Academy of Sciences, Moscow 119991, Russia}

\author{V. Zolotarevskiy}
\affiliation{
Faculty of Mechanical Engineering, Technion -- Israel Institute of Technology, Haifa 32000, Israel}

\author{O. V. Gendelman}
\email{ovgend@tx.technion.ac.il}
\affiliation{
Faculty of Mechanical Engineering, Technion -- Israel Institute of Technology,
Haifa 32000, Israel}

\date{\today}

\begin{abstract}
The paper considers heat transport in diatomic one-dimensional lattices, containing equal amounts
of particles with different masses. Ordering of the particles in the chain is governed
by single correlation parameter -- the probability for two neighboring particles to have the same
mass. As this parameter grows from zero to unity, the structure of the chain varies from regular
staggering chain to completely random configuration, and then -- to very long clusters of particles with equal masses.
Therefore, this correlation parameter allows a control of typical cluster size in the chain.
In order to explore different regimes of the heat transport, two interatomic potentials are considered.
The first one is an infinite potential wall, corresponding to instantaneous elastic collisions
between the neighboring particles. In homogeneous chains such interaction leads to an anomalous
heat transport. The other one is classical Lennard-Jones interatomic potential, which leads
to a normal heat transport. The simulations demonstrate that the correlated disorder of the
particle arrangement does not change the convergence properties of the heat conduction coefficient,
but essentially modifies its value. For the collision potential, one observes essential growth
of the coefficient for fixed chain length as the limit of large homogeneous clusters is approached.
The thermal transport in these models remains superdiffusive. In the Lennard-Jones chain the effect
of correlation appears to be not monotonous in the limit of low temperatures. This behavior stems
from the competition between formation of long clusters mentioned above, and Anderson localization
close to the staggering ordered state.

\end{abstract}

\maketitle

\section{Introduction}
Heat conductivity in one-dimensional (1D) lattices is a
well-known classical problem related to the microscopic
foundation of Fourier's law. The problem started from the
famous work of Fermi, Pasta, and Ulam (FPU) \cite{FPU}, where
an abnormal process of heat transfer was initially revealed.
Numerous aspect of the problem were widely addressed over last two decades \cite{LLP03,LLP08,Dhar08}.
It was established that mere nonlinearity of the interparticle interactions in one-dimensional models
is insufficient for convergence of the heat conduction coefficient in the thermodynamical limit.
Recently it was stated that if the potential of interaction is bounded (allows dissociation of the
neighboring particles), then the heat conductivity converges \cite{SK14,GS14}. In the same time,
it is still not clear whether this condition is necessary and what is relationship between necessary
and sufficient requirements for the interatomic potential, and dimensionality of the problem.

In this paper we address other system properties, which directly effect the heat transport --
a homogeneity and an ordering in the lattice.
Modifications of the heat transport caused by inhomogeneities in the mass or potential
distribution were explored starting from the pioneering work on harmonic system with different
masses \cite{CL71}. Later works considered isotopic disorder \cite{OL74}, harmonic \cite{V79}
and annharmonic random chains \cite{BM90,DL08}.
However, main attention of the studies on the nonhomogeneous systems has been attributed to chains
with staggering \cite{Hatano, diatomic} or randomly distributed masses.
In particular, numerous papers simulated the
heat transport in one-dimensional diatomic hard-point gas. This system is especially interesting,
since the case of equal masses corresponds to completely integrable linear homogeneous billiard
(even its version with external on-site potential is integrable, see, e.g. \cite{GS04}).
So, the staggering masses constitute a perturbation of the integrable case, and the  question is
whether this perturbation will lead to normal heat transport. Numerous  numeric  works on the
diatomic billiard with staggering particles led to a conclusion that the heat conduction
coefficient  $\kappa$  in this system diverges in the thermodynamical limit as $\kappa \sim N^\alpha$
for certain $\alpha>0$ \cite{Dhar01,STZ02, GNY02,CP03}.
More detailed recent exploration of this system \cite{cwcb14} demonstrated,
that when the mass ratio is slightly different from one, it is not possible to exclude
normal heat conduction over longer and longer sizes as the integrable
limit is approached. In the same time, it was conjectured \cite{cwcb14} that also in this case
the heat conductivity diverges, but for longer chains than are available for simulation.

In this paper we also consider the thermal transport in the chain models, which include the
particles with two different masses, but do that in more general setting. Key difference from the
previous studies is that the ordering of these particles in the chains is neither perfectly
periodic (homogeneous or staggering) nor completely random. Instead, the mass distribution in the
chain is characterized by the correlated disorder. More exactly, the masses of neighboring
particles are equal with probability $p_m$, and $0 \le p_m \le 1$. Increase of this parameter
favors clustering in the chain. All previously studied cases (complete order and uncorrelated disorder)
correspond to special values of $p_m$. The effect of the correlated disorder is studied
for the system of rigid particles with abnormal heat conductivity, and also for a model with
convergent heat conductivity -- Lennard-Jones (LJ) chain.
It is demonstrated that in any of these systems the correlated disorder does not modify the convergence
properties of the heat conduction coefficient in the thermodynamical limit, but strongly effected
the quantitative characteristics of the heat transport.

\section{Diatomic gas of rigid particles with random mass distribution}

Let us consider the one-dimensional chain of rigid rods with size $d>0$, with masses which can
take only two values $M_1=1$ and $M_2=m\ge 1$ (dimensionless parameter $m$ is interpreted as
the mass ratio). To be specific, we choose the length of the rods as $d=0.1$ and average distance
between their centers $a=1$. In other terms, numeric density of the rods is adopted to be unit.
The rods are numbered in ascending order of coordinates of their centers.
Hamiltonian of the system is expressed in the following form:
\begin{equation}
H=\sum_n[\frac{p_n^2}{2m_n}+V(x_{n+1}-x_n)].
\label{f1}
\end{equation}
Here $m_n$ is a mass of the   $n$-th rod, $x_n$ -- coordinate of its center, $p_n=m_n\dot{x}_n$.
The interaction of absolutely rigid particles is described  by the following
hard-core potential:
\begin{equation}
V(r)=\infty~~\mbox{if}~~r\le d,~~~V(r)=0~~\mbox{if}~~r>d.
\label{f2}
\end{equation}
Potential function (\ref{f2}) describes instantaneous elastic collisions of the rods.
In this one-dimensional chain, only neighboring rods can collide, and collision of the rod
number $n$ with the rod number $n+1$ occurs if the distance between their centers $x_{n+1}-x_{n}=d$.
As it is well-known, if before the collision velocities of the rods are $v_n=\dot{x}_n$ and $v_{n+1}=\dot{x}_{n+1}$,
then after the collision the velocities will take the following values:
\begin{eqnarray}
v'_n &=& [2m_{n+1}v_{n+1}+(m_n-m_{n+1})v_n]/(m_n+m_{n+1})\nonumber\\
v'_{n+1} &=& [2m_nv_n+(m_{n+1}-m_n)v_{n+1}]/(m_n+m_{n+1}).\nonumber
\end{eqnarray}
Between the collisions, the rods move as free particles.

In considered model, each rod can have the mass $m_n=1$ or the mass $m_n=m$ with equal concentration.
To describe the correlation between the masses of neighboring rods, we define additional parameter
$0\le p_m \le 1$, which denotes a probability for the neighboring rod to have the same mass.
The case  $p_m=0$ corresponds to the chain with alternating masses, the case $p_m=0.5$ corresponds
to completely random distribution of the masses in the chain (lack of correlation between the neighbors).
The higher value of $p_m$, the longer clusters of particles with equal equal masses are expected
in the chain. Average length of such homogeneous clusters is estimated as
$N_p \sim 1/(1-p_m)$. In the limit $p_m \longrightarrow 1$ the chain becomes almost homogeneous,
in other terms, it contains the homogeneous clusters of diverging length $N_p \longrightarrow \infty$).

For simulation of the heat transport in this model with $N$ rods, we include the interaction of
terminal rods with boundary thermostats. The rod with $n=1$ interacts with thermostat of
temperature $T_+$, the rod number $N$ -- with thermostat of temperature $T_-$.
Interaction of the first rod with the thermostat occurs when $x_1=d/2$. In this moment,
the velocity of this rod is re-ascribed to $v_1>0$; the latter is random with Maxwell distribution à
$$
P(v)=(|v|m/T)\exp(-v^2m/2T)
$$
with mass $m=m_1$ and temperature $T=T_+$. Similarly, the rod $N$ interacts with Maxwell
thermostat when $x_N=N-d/2$. In this moment, the velocity of this
is re-ascribed to random value $v_N<0$ with Maxwell distribution with mass $m=m_N$ and temperature $T=T_-$.

As it was mentioned above, the rods interact with the thermostats only when they collide with the boundaries.
In the moment of collision $t=t_j$ the thermostat changes the energy of the terminal rod by value
$\Delta E_i(t_j)=m_i[v_i^2(t_j+0)-v_i^2(t_j-0)]/2$,
where $i=1,N$. If over time interval $[0,t]$ there were $N_t$ collisions of the terminal rod with
the boundary in time instances ($\{t_j\}_{j=1}^{N_t}\in [0,t]$), then the average work done by
the thermostat is expressed as
$$
j_i(t)=\frac{1}{t}\sum_{j=1}^{N_t}\Delta E_i(t_j).
$$
and its average power $J_i=\lim_{t\rightarrow \infty}j_i(t)$.

For simulation of the heat transport in the system we choose the following initial conditions:
$$
x_n(0)=n-1/2,~~\dot{x}_n(0)=v_n,~~n=1,2,...,N,
$$
Here $v_n$ is a random value with Maxwell distribution à
$P(v)=\sqrt{m_n/2\pi T}\exp[-m_nv^2/2T]$, where $T=(T_++T_-)/2$.

Following paper \cite{cwcb14}, the temperature of the left boundary is set to $T_+=6$,
and of the right boundary -- to $T_-=4$. Then, long time dynamics of the system is simulated.
It should be mentioned that this dynamics does not depend on the absolute values of
the temperatures, but only on the ratio $T_+/T_-$.
After initial transient and formation of stationary heat flux, average powers
of the thermostats $J_1$, $J_N$ are computed. The heat flux in the system should satisfy relationship
$J=J_1=-J_N$, which can serve also as a test for validity of the numeric procedure.
In the simulations these equations were obeyed with very high accuracy.
Distribution of the local temperature in the chain is determined as $T_n=\langle \dot{x}_n^2(t)\rangle_t$.
\begin{figure}[tb]
\includegraphics[angle=0, width=1\linewidth]{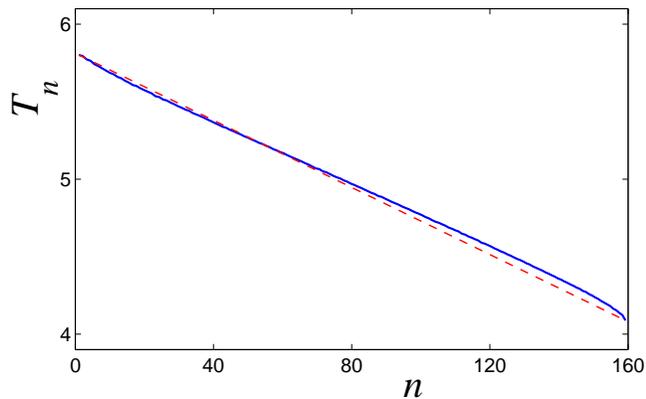}
\caption{
(Color online)
Distribution of temperature $T_n$ along the chain of length
$N=160$. The mass ratio is $m=1.4$, parameter of correlation $p_m=0$.
The dashed line represents the linear temperature gradient, used
for computation of the heat conduction coefficient $\kappa(N)$.}
\label{fig01}
\end{figure}

Simulation of the heat transport demonstrates that the collisions of the rods with different mass
bring about formation of a linear thermal gradient in the chain, see Figure \ref{fig01}.
The effect of thermal resistance reveals itself at the ends of the chain.
As a result of this effect it
turns out that the temperature of the leftmost rod $T_1<T_+$, and of the rightmost rod $T_N>T_-$.
This discrepancy decreases with an increase of the chain length. To avoid ambiguities related
to this thermal resistance, value of the heat conduction coefficient was calculated
in accordance with actual temperatures of the terminal rods with the following formula:
\begin{equation}
\kappa(N)=JN/(T_1-T_N). \label{f3}
\end{equation}

\begin{figure}[tb]
\includegraphics[angle=0, width=1\linewidth]{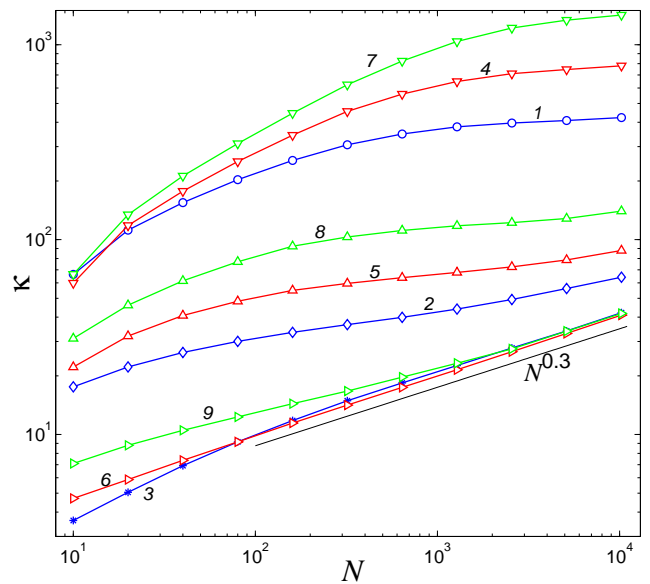}
\caption{
(Color online)
Dependence of the heat conduction coefficient $\kappa$ on the length of chain $N$
for mass ratios $m=1.1$, 1.4, 3 and for three values of the mass correlation parameter:
$p_m=0$ (blue curves 1, 2, 3),  $p_m=0.5$ (red curves 4, 5, 6), $p_m=0.75$
(green curves 7, 8, 9).
}
\label{fig02}
\end{figure}

Dependence of the heat conduction coefficient $\kappa$ on the chain length $N$ for different
values of the mass ratio $m$ and mass correlation parameter $p_m$ is presented in Figure \ref{fig02}.
One can learn from this Figure, that for the case $p_m=0$ (perfectly ordered diatomic chain)
the dependence $\kappa(N)$ is similar to results presented previously in paper \cite{cwcb14}.
For large value of the mass ratio $m=3$ the heat conduction coefficient increases approximately as $N^{0.3}$.
However, for small mass ratio $m=1.1$ for $N\sim 10^4$ the growth of the heat conduction
coefficient slows down. It is extremely difficult to check the behavior of the coefficient
for essentially longer chains.
\begin{figure}[tb]
\includegraphics[angle=0, width=1\linewidth]{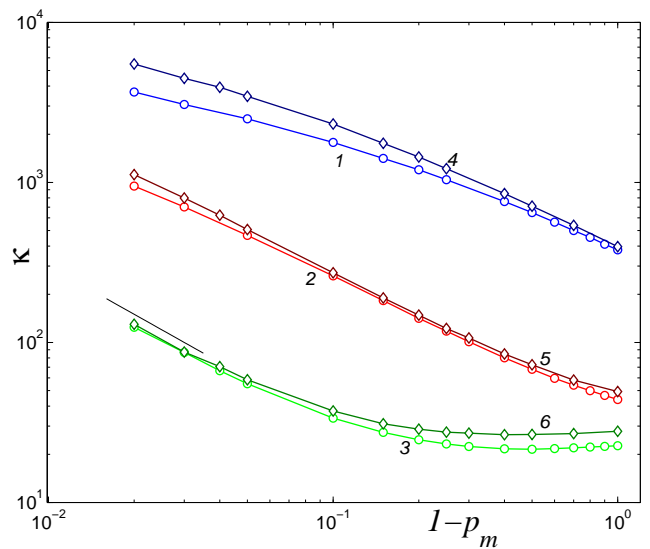}
\caption{
(Color online)
Dependence of the heat conduction coefficient $\kappa$
on the mass correlation parameter $p_m$ for chain length $N=1280$ (curves 1, 2, 3),
$N=2560$ (curves 4, 5,6) and mass ratios $m=1.1$, 1.4, 3.
Straight line corresponds to expression $\kappa=3/(1-p_m)$.
}
\label{fig03}
\end{figure}

Increase of the mass correlation parameter $p_m$ for fixed mass ratio and chain length leads
to an increase of $\kappa$, as it is demonstrated in Figures \ref{fig02}, \ref{fig03}.
In the limit $p_m\longrightarrow 1$ one obtains $\kappa(N)\longrightarrow\infty$. The reason is that
the modification of momenta of the individual rods is possible if the colliding
particles have different masses. Average length of clusters of the rods with the same mass $N_p$
grows together with the mass correlation parameter $p_m$, and relative number of the scattering
events collisions of the rods with different mass decreases. So, one can expect that
for large $N$ the heat conduction coefficient  $\kappa$ will be proportional to
the average length  of the cluster with equal masses, which can be estimated as $N_p \sim 1/(1-p_m)$.
As it follows from Figure \ref{fig03}, in the case of large mass ratio $m=3$ we observe the
divergence of the heat conduction coefficient $\kappa \sim 1/(1-p_m)$. For smaller values
of the mass ratio, no clear power law is observed. For $m=1.4$ it is still possible to argue
approaching the power law, but in the case of $m=1.1$ the crossover is much slower,
and simulation of longer chains is required.

The convergence behavior of $\kappa$ with increase of $p_m$ strongly resembles
the situation with $m\longrightarrow 1$.

As it was mentioned before, our results for $p_m=0$ reproduce simulations presented in \cite{cwcb14}.
There the authors have mentioned that on the basis of the simulation results it is not possible
to reject possibility of convergence of the heat conduction coefficient for the case of small mass
ratios. From Figure \ref{fig02} one can observe that for larger values of the mass correlation
parameter the situation remains qualitatively the same. The authors of \cite{cwcb14} also
conjectured that the observed "convergence-like" behavior is transient, and the heat conduction
coefficient should attain the divergence regime for higher values of $N$. Results presented
in Figure \ref{fig03} give a clue that the divergence of $\kappa$ in the limit of
$p_m \longrightarrow 1$ exhibits similar very long crossovers. In the next Section, we attempt
to achieve numeric evidence on divergence of the heat conduction coefficient in the "crossover"
situations  by simulating the dynamics of temperature spreading in the chain.
\begin{figure}[tb]
\includegraphics[angle=0, width=1\linewidth]{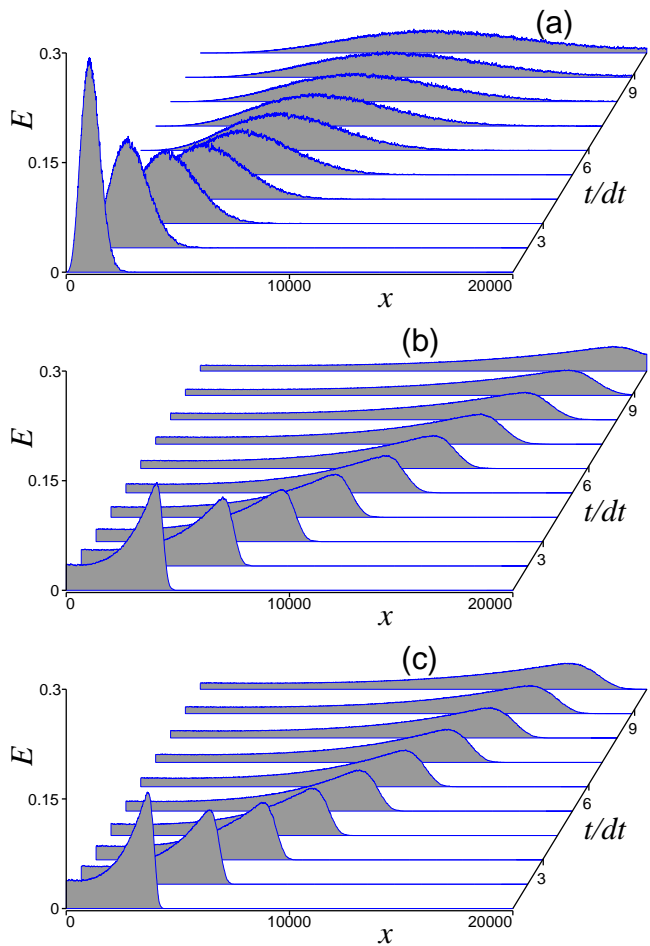}
\caption{
Transport of initial energy pulse in the chain with $N=20000$ particles after initial
thermal excitation at the left end.
(a) $m=1$, time step $dt=144$; (b) $m=1.4$, $dt=3600$; (c) $m=3$, $dt=3960$, $p_m=0$.
Energy distribution in the chain $E(x)$ is plotted for multiples of $dt$.
}
\label{fig04}
\end{figure}
%
\section{Diffusion of energy pulses in the chain with random mass distribution}

For simulation of the energy diffusion the chain comprising $N=20000$ particles (rods) is considered.
The thermostats are removed and the terminal rods elastically interact with the walls.
Initially all the rods are randomly placed in the interval $(d, N-d)$ and 60 rods at the left end
of the system are thermalized with average temperature $T_{in}$. The rest of the rods (for $n>60$)
are thermalized with relatively small average temperature $T_0 < T_{in}$.
Then, the diffusion of energy along the chain is simulated. To improve the accuracy,
all results of such simulations
are averaged over $10^4$ independent realizations of the initial temperature distribution.
For every time instance one obtains energy distribution in the chain
$$
E_n(t)=\sum_{i: x_i\in(n-1,n]}m_i\dot{x}_i^2(t)/2.
$$

Let us at first consider energy diffusion in initially non-thermalized chain.
For this matter we adopt the value $T_{in}=10$ as the initial temperature
of the left end of the chain, and the temperature of the rest part of the chain is $T_0=0$.
Results of these simulations are presented in Figure \ref{fig04}.
The character of the thermal diffusion depends on the value of mass ratio $m$.
For $m=1$ one observes completely ballistic flow -- all energy leaves
the left end of the chain, and the heat pulse becomes wider and lower, since the initial pulse
contains particles with different velocities. For $m>1$ the flow includes both ballistic
and diffusive components, and some residual energy remains at the left end of the system.
\begin{figure}[tb]
\includegraphics[angle=0, width=1\linewidth]{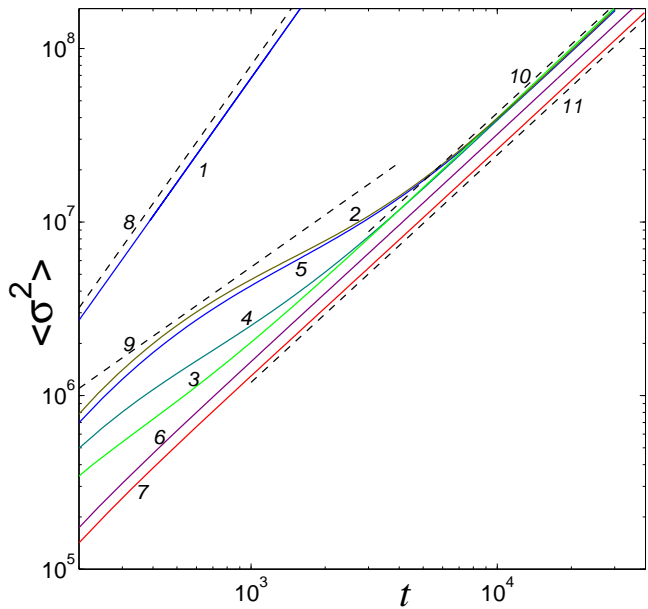}
\caption{
(Color online)
Time dependence of the measure of energy spreading over the chain $\sigma^2$ (double logarithmic scale)
for the set of parameters: $m=1$ (curve 1); $m=1.05$ and $p_m=0$ (curve 2);
$m=1.1$ and $p_m=0$, 0.5, 0.75 (curves 3, 4, 5 respectively); $p_m=0$ and $m=2$, 3 (curves 6, 7 respectively).
Dashed lines correspond to power laws $t^\beta$ for the values of exponents
$\beta=2$, 1 and 1.31 (curves 8, 9 and 10, 11 respectively).
The temperature of the chain is $T_0=0$.
}
\label{fig06}
\end{figure}

It is common to describe evolution of the temperature distribution in the chain with the help
of the following "mean quadratic"\
deflection from the initial state (concentration at the left end
of the chain):
\begin{equation}
\sigma^2=\sum_{n=1}^N(n-1/2)^2e_n. \label{f4}
\end{equation}
Here $\{e_n=(E_n-E_0)/2E\}_{n=1}^N$ characterizes the energy distribution in given time instant,
$E=\sum_{n=1}^N(E_n-E_0)$ is a total constant energy of the chain, $E_0=T_0/2$
-- energy level of initially "cold" part of the chain.

Numeric simulation demonstrates that for non-thermalized chain ($T_0=0)$
the mean quadratic deflection at large times asymptotically
obeys power law: $\langle\sigma^2\rangle\sim t^\beta$ when $t\nearrow\infty$.
These results are illustrated in Figure \ref{fig06}.

The works \cite{dku03prl,cdp05prl,dsd13pre,lhlrl14prl} argue that in systems with collision dynamics,
where the particles undergo L\'evy flights, the exponent $\beta$, which describes the mean square
displacement of particles $\langle x^2(t)\rangle\sim t^\beta$, should be related with $\alpha$
through equation
\begin{equation}
\alpha=\beta-1. \label{f5}
\end{equation}
This equation implies that normal diffusion ($\beta=1$) leads to the normal (non-divergent)
heat conductivity ($\alpha= 0$). In the same time, superdiffusion ($\beta >1$) corresponds
to divergent ($\alpha >0$) heat conductivity.

Numeric results in Figure \ref{fig06} for homogeneous chain with $m=1$ yield value $\beta=2$,
which corresponds to purely ballistic energy transport ($\alpha=1$).
For mass ratios $m=1.05$, 1.1, 2 and 3 we obtain $\beta=1.31$, which is related to the
superdiffusion of energy in non-thermalized diatomic chain with  mass ratio $m>1$.
The correlation in mass distribution along the chain does not lead to a change in the type
of diffusion. So, for $m=1.1$ the value of the exponent $\beta$ does not depend on the value $p_m<1$.
For $m=3$ the measured exponent of divergence of the heat conductivity is $\alpha=0.3$,
such that for all the values $p_m$ there is a good agreement with the relation (\ref{f5}).
\begin{figure}[tb]
\includegraphics[angle=0, width=1\linewidth]{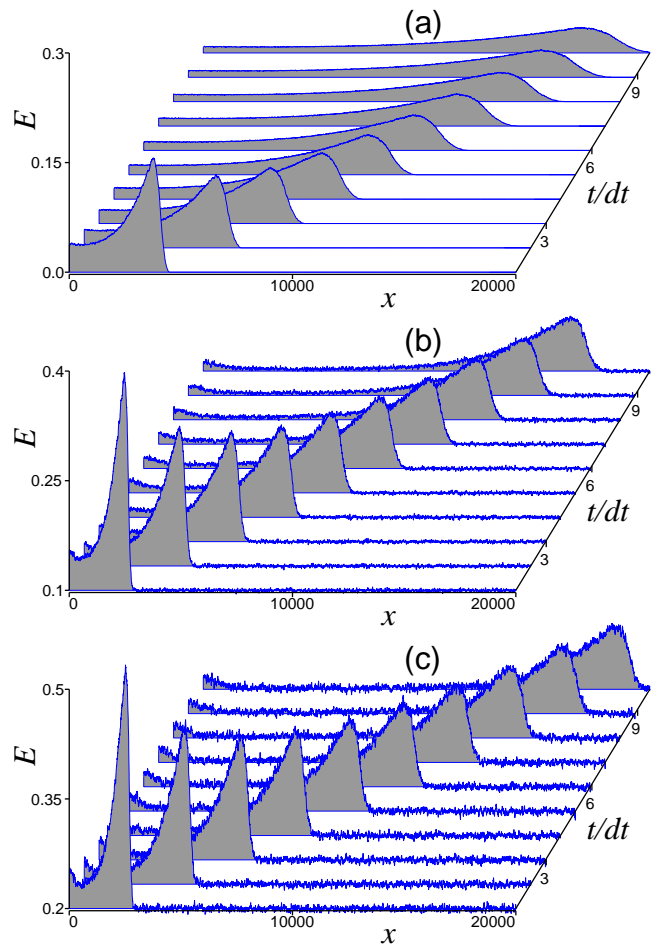}
\caption{
Transport of initial energy pulse in the chain with $N=20000$ particles after initial thermal
excitation at the left end $T_{in}=10+T_0$ for temperature of the chain (a) $T_0=0$,
time step $dt=3600$; (b) $T_0=0.2$, $dt=1500$ and (c) $T_0=0.4$, $dt=1320$.
Mass correlation parameter is $p_m=0$, mass ratio $m=2$.
Energy distribution in the chain $E(x)$ is plotted for multiples of $dt$.
}
\label{fig07}
\end{figure}
\begin{figure}[tb]
\includegraphics[angle=0, width=1\linewidth]{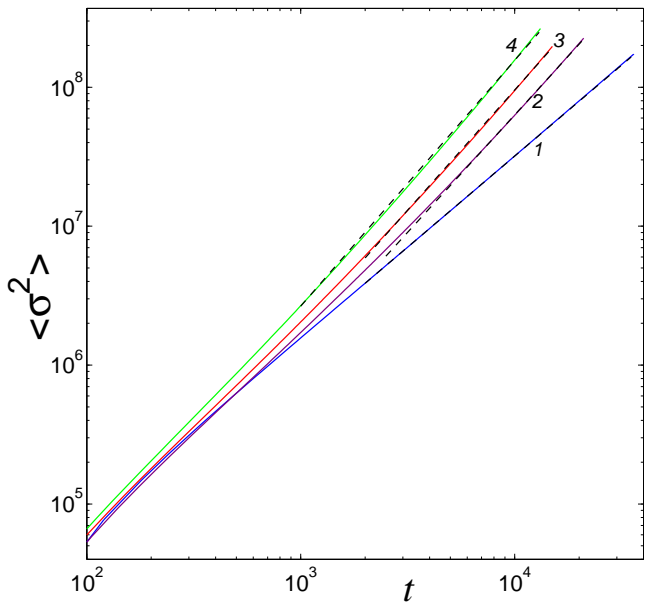}
\caption{
(Color online)
Time dependence of the measure of energy spreading over the chain $\sigma^2$ (double logarithmic scale)
for temperature of the chain $T_0=0$, 0.1, 0.2 and 0.4 (curve 1, 2, 3 and 4).
Mass ratio $m=2$, mass correlation parameter $p_m=0$.
Dashed straight lines correspond to power laws $t^\beta$ for the values of exponents
$\beta=1.31$, 1.69, 1.73 and 1.77 (curves 1, 2, 3 and  4 respectively).
}
\label{fig08}
\end{figure}

\begin{figure}[tb]
\includegraphics[angle=0, width=1\linewidth]{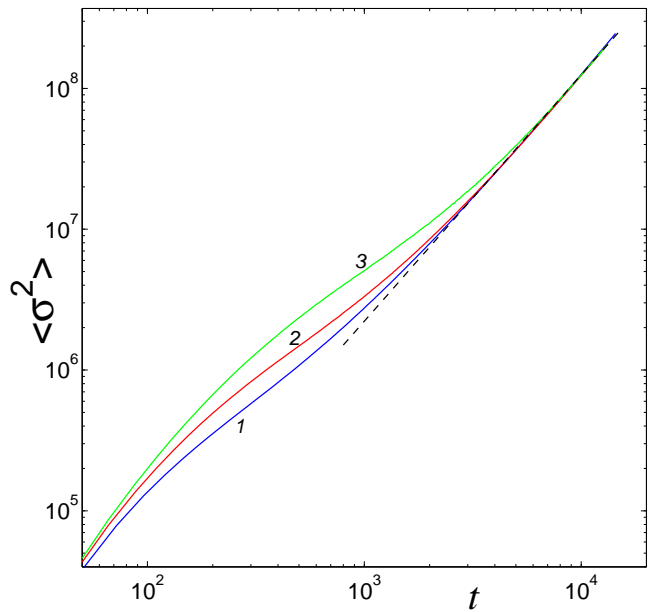}
\caption{
(Color online)
Time dependence of the measure of energy spreading over the chain $\sigma^2$ (double logarithmic scale)
for the chain with mass correlation parameter $p_m=0$, 0.5 and 0.75 (curve 1, 2 and 3).
Mass ratio $m=1.1$, temperature of the chain $T_0=0.2$.
Dashed straight line corresponds to power law $t^{1.75}$.
}
\label{fig09}
\end{figure}

Let us consider now diffusion of energy in a thermalized chain  with  $T_0>0$. Figure \ref{fig07}
illustrates that dynamics of the energy pulse significantly depends on the temperature
of the chain. An increase of the temperature $T_0$ leads to an increase in the rate of propagation
of the heat pulse along the chain. The exponent of the energy diffusion $\beta$ increases with
an increase in the mean temperature $T_0$  -- see Fig. \ref{fig08}. So, in the diatomic
chain with staggering masses ($p_m=0$) for mass ratio $m=2$ with $T_0=0$ we obtain the energy
diffusion exponent $\beta=1.31$, for $T_0=0.1$ $\beta=1.69$, for $T_0=0.2$, $\beta=1.73$,
and for $T_0=1.77$ the value of the exponent $\beta=1.77$. We are not aware of any significant
dependence of $\alpha$ on average temperature (or, in this collisional model, on ratio of the
temperatures on the cold and the hot ends of the chain). Thus, the obtained results
suggest that the relationship between $\alpha$ and $\beta$ (\ref{f5}) is satisfied only for $T_0=0$. Both in the cases $T_0>0$ and $T_0=0$ the value of the
diffusion exponent $\beta$ does not exhibit any visible dependence on the correlation
of the mass distribution of the rods in the chain. The correlation parameter $p_m$ only effects
the time at which the system achieves the regime of steady superdiffusion -- see Fig. \ref{fig09}.

The results of the numerical modeling in a random diatomic chain imply that for mass ratio $m>1$
and for any correlation of the mass distribution of the chain, i.e. for any value of the
parameter $0\le p_m<1$, the heat transport in the diatomic chain is superdiffusive for all studied
values of mass ratio and correlation parameter. These results include also the cases, in which it
was not possible to get clear evidence of divergence of the heat conduction coefficient
in the equilibrium simulations.

\section{Heat transport in random diatomic Lennard-Jones chain}

To explore the effect of mass clustering in the diatomic lattice with convergent heat
conductivity, we consider a chain with common Lennard-Jones (LJ) interatomic potential:
\begin{equation}
V(r)=4\epsilon[(\sigma/r)^6-1/2]^2. \label{f6}
\end{equation}
The following values of parameters are used: $\epsilon=1/72$, $\sigma=2^{-1/6}$.
For this choice of the parameters, the potential has a minimum for the distance between particles
$r_0=1$ and stiffness in the minimum $k=V''(r_0)=1$. Consequently, the sound velocity
in the homogeneous ($m=1$) chain $v_0=r_0\sqrt{k/m}=1$.

Recently it was established that the homogeneous Lennard-Jones chain has convergent heat
conductivity \cite{SK14,GS14}.
Efficient phonon scattering exists due to thermally activated large elongations of
the interparticle bonds; possibility of such large elongations stems from boundedness
of the LJ potential \cite{GS14}. All simulations described below confirm the convergence of
the heat conduction in this system, although this issue is still debated \cite{NJP}.
One might expect that inhomogeneities in the chain enhance the phonon scattering and,
as a consequence, decrease the heat conductivity. Therefore for $m>1$ the heat conductivity
of the LJ chain should be convergent, as in the homogeneous case.

To check this idea, the heat transport in random LJ chain with end regions attached to Langevin
thermostats is simulated. The method of simulation is described in details in paper \cite{SK14}.
The LJ chain contains $N=80+10\times 2^l$, $l=1,2,...,10$  particles
with fixed boundary conditions. Thee leftmost 40 particles are attached to the Langevin thermostat
with temperature $T_+$ and 40 rightmost particles -- to the Langevin thermostat with temperature
$T_-$, where $T_\pm=(1\pm 0.1)T$. Here $T$ is the average temperature of the chain.
Local temperature distribution in the chain is defined as $T_n=\langle \dot{x}_n^2\rangle$, à
the heat flux -- as $J=\langle J_n\rangle$, where $J_n=-\dot{x}_n V'(x_{n+1}-x_n)$
is a local instantaneous flux of energy through site number $n$.
In internal fragment of the chain $40<n\le N-40$ one observes constant heat flux and approximately
linear thermal profile. Thus, it is possible to evaluate the heat conduction coefficient
as $\kappa(N_i)=JN_i/(T_{41}-T_{N-40})$, $N_i=N-81$
is the length of the chain fragment between the thermostats.
\begin{figure}[tb]
\includegraphics[angle=0, width=1\linewidth]{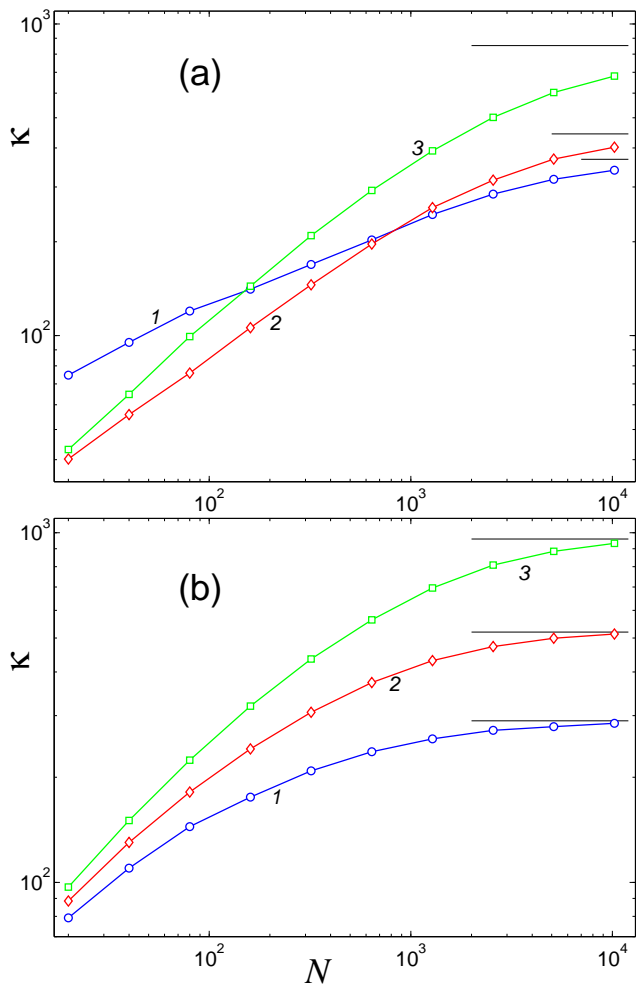}
\caption{
(Color online)
Dependence of the thermal conductivity $\kappa$ on the length $N$ of the diatomic LJ chain
for temperature (a) $T=0.0005$ and (b) $T=0.2$. The mass ratio $m=1.1$.
The curves represent the results for three values of the mass correlation parameter
$p_m=0$, 0.5 and 0.75 (curves 1, 2 and 3 respectively).
Horizontal lines correspond to the values of the heat conduction coefficient
obtained with the help of Green-Kubo relation.
}
\label{fig11}
\end{figure}

Thermal conductivity can also be found with the help of Green-Kubo relation \cite{kubo}
\begin{equation}
\kappa=\lim_{t\rightarrow\infty}\lim_{N\rightarrow\infty}\frac{1}{NT^2}\int_0^tc(\tau)d\tau,
\label{f7}
\end{equation}
where $c(t)=\langle J_s(\tau)J_s(\tau-t)\rangle_\tau$  autocorrelation function of the heat flux;
the latter is defined as $J_s(t)=\sum_nJ_n(t)$.

To find the autocorrelation function $c(t)$, we simulate a cyclic chain with $N=10^4$
units, initially completely immersed in the Langevin thermostat with temperature $T$.
After thermal equilibration with the thermostat, the latter is removed and Hamiltonian dynamics
of the chain is simulated. To increase the accuracy of the autocorrelation function measurement,
its value is averaged over $10^4$ independent realizations of the initial chain thermalization.

In the case of small temperatures $T\ll \epsilon$ the heat transport occurs due to propagation
of weakly interacting linear waves (as it is commonly accepted, we use the term "phonons" despite
purely classical character of our model). If the temperatures are large, $T\gg \epsilon$ only hard
cores of the interatomic potentials have significant effect on the heat transport.
So, the LJ chain can be considered as a set of almost completely rigid particles, and
the heat transport occurs due to the collisions between these particles. Since the heat transport
mechanisms for high and low temperatures are rather different, it is natural to expect different
effect of the random mass distribution in these two cases.

First, let us consider the heat transport in the LJ chain for relatively low temperature $T=0.0005$.
In this limit the system becomes very close to the chain with parabolic interatomic potential.
Therefore, the diatomic LJ chain with staggering masses (mass correlation parameter $p_m=0$)
becomes close to well-known homogeneous diatomic chain with its typical acoustical
and optical phonon branches. In the same time, for random chain with mass correlation
parameter $0<p_m<1$ it is well-known that some vibration modes become localized
(Anderson localization). So, one can expect that as $p_m$ grows above zero and
the homogeneous diatomic chain becomes randomized, the heat conductivity of the chain
should decrease. In the same time, as $p_m$ approaches unity, the chain becomes closer
to the monoatomic case, and the heat conductivity should increase.
Therefore, contrary to the chain of rigid particles explored above, one can
qualitatively predict non-monotonous dependence of the heat conduction
coefficient on the mass correlation parameter $p_m$ for fixed temperature and mass ration.
Moreover, it is anticipated that this effect will be more pronounced for higher mass ratio.

Dependence of the heat conduction coefficient $\kappa$ on the chain length $N$ for different
values of $p_m$ is presented in Fig. \ref{fig11}.
For the mass ratio close to unity $m=1.1$, the increase of $p_m$ leads
to relatively mild decrease of the heat conductivity only for short chains with $N<10^3$,
as it is illustrated in  Figure \ref{fig11} (a). Possible explanation is that the chain
is rather close to the homogeneity, thus the Anderson localization is less pronounced
and other effects (such as grow of expected length of the homogeneous clusters) can
compete and blur the effect of localization. For longer chains, the heat transport
is governed by phonons with large wavelength, which are less sensitive
to small inhomogeneities of the mass distribution.
\begin{figure}[tb]
\includegraphics[angle=0, width=1\linewidth]{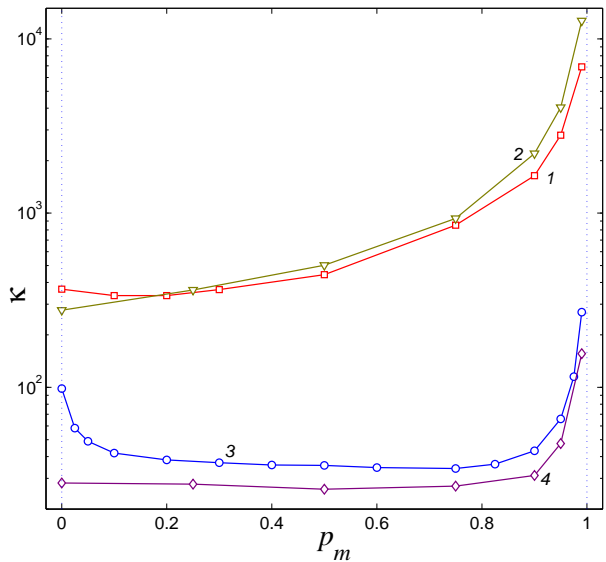}
\caption{
(Color online)
Dependence of the thermal conductivity $\kappa$
on mass correlation coefficient $p_m$ in the LJ chain  with mass ratio $m=1.1$
and temperatures $T=0.0005$ and $T=0.2$ (curves 1 and 2)
and for $m=2$, $T=0.0005$ and $T=0.01$ (curves 3 and 4).
}
\label{fig12}
\end{figure}

For high temperature $T=0.2$ the chain dynamics is primarily governed by collisions between
hard repulsive cores of the particles. It is possible to say that in this case the LJ chain
is close 1D diatomic gas model, but not completely equivalent to it. The crucial difference
is that for any finite values of the temperature the time of collisions is finite. Therefore,
contrary to the 1D gas considered above, triple collisions of the particles have a nonzero
probability. In recent paper \cite{GS14} it is demonstrated that such triple collisions provide
efficient scattering mechanism for energy transport in the system and promote convergence
of the heat conduction coefficient. Consequently, one should expect a monotonous growth
of the heat conductivity with increasing $p_m$ similarly to the diatomic gas.
Numeric simulation presented in Figure~\ref{fig11} (b) confirms this similarity
(cf. Figure~\ref{fig03}). Important difference is that in the LJ chain the heat
conductivity for $p_m\longrightarrow 1$ remains convergent in the thermodynamical limit.

Due to this convergence, it is reasonable to consider a dependence of the limit value of
the heat conduction coefficient on the mass correlation parameter $p_m$. This dependence has been analyzed,
for the chain with $N=20000$ particles for different temperatures and mass ratios with the help
of Green-Kubo relation (\ref{f7}). Methodology of this simulation has been described above,
and the results are presented in Figure~\ref{fig12}.

From Figure~\ref{fig12} one can learn that the character of dependence $\kappa(p_m)$ substantially
depends on the temperature and mass ratio. Common feature of all these dependencies is a
rapid growth of the heat conductivity as $p_m\longrightarrow 1$.
For relatively low temperature $T=0.0005$ and for both considered mass ration $m=1.1$
and $m=2$ these dependencies are not monotonous. In particular, for $m=1.1$ the heat conductivity
achieves a minimum for $p_m=0.2$ and for $m=2$ the minimum is achieved for $p_m=0.75$
(see Fig.~\ref{fig12}, curves 1 and 3). As it was mentioned above, this non-monotonicity
may be attributed to Anderson localization of oscillatory states in the chain with random
structure -- two terminal states with $p_m=0$ and $p_m=1$ are ordered.
This effect is much more pronounced for higher mass ratio $m=2$.

For the case of relatively high temperatures this phenomenon of non-monotonicity disappears.
Namely, for mass ratio $m=2$ and temperature $T=0.01$ the heat conductivity is almost constant
over the interval $0\le p_m<0.9$ and rapidly grows for $p_m>0.9$ (see Figure~\ref{fig12}, curve 4).
For mass ratio $m=1.1$ and temperature $T=0.2$ the heat conductivity exhibits a monotonous growth
over the whole interval  $0<p_m<1$ (see Figure~\ref{fig12}, curve 2). This behavior is another
illustration of primarily collisional dynamics of the LJ chain in the limit of high temperatures
-- it behaves similarly to the gas of rigid particles, cf. Figure~\ref{fig03}.
It is important to observe that in this regime there is no significant physical difference
between the ordered staggered state with $p_m=0$ and disordered states with small values of $p_m$.
The reason is that the collision mechanism is local and not sensitive to the short-range order or disorder.

\section{Concluding remarks}

The first conclusion to mention here is that the internal correlations in the particle ordering
do not modify the convergence properties of the heat conduction coefficient. In the model of
colliding rods, superdiffusive transport has been detected for all explored values of the mass
ratio an the correlation coefficient, including those for which direct equilibrium simulation
of the heat transport is beyond current computational capabilities. It is worth while mentioning,
that the superdiffusion exponent turned out to be dependent on initial heating
of the chain beyond the initial thermal pulse.

In the same time, the correlations in the mass disorder have significant effect on the value
of the heat conductivity. This effect reveals itself through two competing factors -- growth
of the clusters of particles with the same mass, and frustration of the perfect disordered state.
In the case of purely collisional dynamics, the latter factor plays no role, and the heat
conductivity of the chain with given length monotonously increases with increase of the
correlation parameter. In the case of relatively small temperatures in LJ chain both factors
are significant, and the dependence is not monotonous due to Anderson localization as the
system departs from the ordered staggering state. For large temperatures, the quantitative
dependence of the heat conduction coefficient on the correlation parameter is similar
to that in the model of colliding rods; however, the thermal transport appears to remain convergent.

\section{Acknowledgments}
The authors are very grateful to Israel Science Foundation (grant 838/13) and to Lady Davis
Fellowship Trust for financial support of their work. A.V.S. is grateful to the Joint
Supercomputer Center of the Russian Academy of Sciences for the use of computer facilities.


\begin{references}
\bibitem{FPU}
E. Fermi, J. Past,a and S. Ulam.
Studies of nonlinear problems.
Los Alamos Report No. LA.1940 (1955).
\bibitem{LLP03}
S. Lepri, R. Livi, and A. Politi.
Phys. Rep. {\bf 377}, 1 (2003).
\bibitem{LLP08}
S. Lepri, R. Livi, and A. Politi.
\emph{Anomalous Transport: Foundations and Applications} Weinheim:Wiley-VCH Verlag, Ch. 10, 2008).
\bibitem{Dhar08}
A. Dhar.
Heat transport in low-dimensional systems.
Advances in Physics  {\bf 57}, 5, 457-537 (2008).
\bibitem{SK14}
A. V. Savin and Yu. A. Kosevich.
Thermal conductivity of molecular chains with asymmetric potentials of pair interactions.
Phys. Rev. E {\bf 89}, 032102 (2014).
%
\bibitem{GS14}
O. V. Gendelman and A. V. Savin.
Normal heat conductivity in chains capable of dissociation.
EPL, {\bf 106} 34004 (2014).

\bibitem{CL71}
A. Casher and J. L. Lebowitz.
Heat Flow in Regular and Disordered Harmonic Chains.
J. Math. Phys., {\bf 12} 1701 (1971).

\bibitem{OL74}
A. J. O'Connor and J. L. Lebowitz.
Heat conduction and sound transmission in isotopically disordered harmonic crystals.
J. Math. Phys., {\bf 15} 692 (1974).

\bibitem{V79}
T. Verheggen.
Transmission Coefficient and Heat Conduction of a
Harmonic Chain with Random Masses:
Asymptotic Estimates on Products of Random Matrices.
Commun. Math. Phys., {\bf 68} 69 (1979).

\bibitem{BM90}
R. Bourbonnais and R. Maynard.
Energy Transport in One- and Two-Dimensional Anharmonic Lattices with Isotopic Disorder.
Phys. Rev. Lett, {\bf 64} 1397 (1990).

\bibitem{DL08}
A. Dhar and J. L. Lebowitz.
Effect of Phonon-Phonon Interactions on Localization.
Phys. Rev. Lett, {\bf 100} 134301 (2008).

\bibitem{Hatano}
T.Hatano.
Heat conduction in the diatomic Toda lattice revisited
Phys. Rev. E, {\bf 59} R1 (1999)

\bibitem{diatomic}
O.V.Gendelman and L.I.Manevich. 
Nonlinear dynamics of the diatomic Toda lattice and the problem of thermal
conductivity of quasi-one-dimensional crystals.
Sov. Phys. JETP {\bf 75} 271 (1992).

\bibitem{GS04}
O.V.Gendelman and A.V.Savin.
Heat Conduction in a One-Dimensional Chain of Hard Disks with Substrate Potential.
Phys. Rev. Lett. {\bf 92} 074301 (2004).
\bibitem{Dhar01}
A. Dhar,
Heat Conduction in a One-Dimensional Gas of Elastically Colliding Particles of Unequal Masses.
Phys. Rev. Lett. {\bf 86}, 3554 (2001).
\bibitem{STZ02}
A. V. Savin, G. P. Tsironis, and A. V. Zolotaryuk.
Heat Conduction in One-Dimensional Systems with Hard-Point Interparticle Interactions.
Phys. Rev. Lett. {\bf 88}, 154301 (2002).
\bibitem{GNY02}
P. Grassberger, W. Nadler, and L. Yang.
Heat Conduction and Entropy Production in a One-Dimensional Hard-Particle Gas.
Phys. Rev. Lett. {\bf 89}, 180601 (2002).
\bibitem{CP03}
G. Casati and T. Prosen.
Anomalous heat conduction in a one-dimensional ideal gas.
Phys. Rev. E {\bf 67}, 015203 (2003).
\bibitem{cwcb14}
S. Chen, J. Wang, G. Casati, and G. Benenti.
Nonintegrability and the Fourier heat conduction law.
Phys. Rev. E {\bf 90}, 032134 (2014).
\bibitem{dku03prl}
S. Denisov, J. Klafter, and M. Urbakh,
Dynamical Heat Channels,
Phys. Rev. Lett. {\bf 91}, 194301 (2003).
\bibitem{cdp05prl}
P. Cipriani, S. Denisov, and A. Politi,
From Anomalous Energy Diffusion to Levy Walks and Heat Conductivity in One-Dimensional Systems,
Phys. Rev. Lett. {\bf 94}, 244301 (2005).
\bibitem{dsd13pre}
A. Dhar, K. Saito, and B. Derrida,
Exact solution of a L\'evy walk model for anomalous heat transport,
Phys. Rev. E {\bf 87}, 010103(R) (2013).
\bibitem{lhlrl14prl}
S. Liu, P. H\"anggi, N. Li, J. Ren, and B. Li,
Anomalous Heat Diffusion,
Phys. Rev. Lett. {\bf 112}, 040601 (2014).

\bibitem{NJP}
Y. Li, S. Liu, N. Li, P. H\"anggi and B, Li
1D momentum-conserving systems: the conundrum of anomalous
versus normal heat transport,
New Journal of Physics, {\bf 17} 043064 (2015).
\bibitem{kubo}
R. Kubo, M. Toda, and N. Hashitsume, in {\it Statistical Physics II},
edited by P. Fulde, Springer Series in Solid-State Sciences
Vol. 31 (Springer, Berlin, 1991).

\end{references}
\end{document}